\documentclass[epj]{svjour}
\usepackage{psfig}
\usepackage{times}
\begin{document}
\title{${\omega}N$ final state interactions
and $\omega$-meson production from heavy-ion collisions}
\author{
G.I. Lykasov$^1$, W. Cassing$^2$, A. Sibirtsev$^2$, M.V. Rzjanin$^1$}
\institute{
{$^1$\small $^1$    Joint Institute for Nuclear Research} \\
{$^1$\small141980  Dubna, Moscow Region, Russia} \\
{$^1$\small $^2$  Institut f\"ur Theoretische Physik }\\
{$^1$\small Heinrich-Buff-Ring 16, D-35392 Giessen, Germany} }

\date{Received: date / Revised version: date}

\abstract{
We calculate the elastic and inelastic ${\omega}N{\to}{\omega}N$,
${\to}{\pi}N$, ${\to}{\rho}N$,  ${\to}\rho{\pi}N$, 
${\to}\pi{\pi}N$, ${\to}{\sigma}N$  reactions within a boson exchange
approximation where the $\omega\rho\pi$ coupling constant and
form factor are fixed by the reaction ${\pi}N{\to}{\omega}N$
in comparison to the experimental data. We find  rather
large ${\omega}N$ cross sections at
low relative momenta of the $\omega$-meson which leads to a
substantial broadening of the $\omega$-meson width in nuclear
matter. The implications of the
${\omega}N$ final state interactions are studied for
$\omega$ production in $^{12}C{+}^{12}C$, $^{40}Ca{+}^{40}Ca$
and $^{58}Ni{+}^{58}Ni$ reactions at about 2$\cdot$A GeV within the
HSD transport approach; the drastic changes
of the transverse mass spectra relative to a general
$m_T$-scaling (for $\pi^0$ and $\eta$ mesons) might be
controlled experimentally by the TAPS Collaboration.}

\authorrunning{G.I. Lykasov et al.}
\titlerunning{${\omega}N$ final state interactions
and $\omega$-meson production from heavy-ion collisions}

\maketitle

\section{Introduction}
The properties of hadrons in a dense and hot nuclear medium
are of fundamental interest with respect to the question of
chiral symmetry restoration in such an environment, where a
new phase of  strongly interacting matter might be
encountered~\cite{Muller,Ko,Brown,PREP}. The properties of
vector mesons here are of particular interest since these can
be studied experimentally via their decay to dileptons.
Whereas the $\rho$-meson spectral function in the medium has
been discussed to a large extent~\cite{Weise4,Rapp,Peters,Friman4}
the properties of the $\omega$-meson in dense matter especially
at finite relative momentum are achieving increasing
interest~\cite{Schoen,Golub1,Golub2,Klingl1,Klingl2,Klingl4,Friman1,Friman2}.
As in case of the $\rho$-meson the $\omega$-meson properties
at low baryon density are dominantly determined by the interactions
with nucleons; real and imaginary parts of the scattering amplitude
then are determined by dispersion relations. It is thus of
fundamental interest to obtain some information about the
${\omega}N$ scattering cross sections which except
for the channel ${\pi}N{\to}{\omega}N$ are not accessable by
experiment.

In this work we address the latter question in a boson -exchange
approach and study the implications from the ${\omega}N$
final state interactions for $\omega$ production in
nucleus-nucleus reactions around 2 A$\cdot$GeV in the
context of present experiments by the
TAPS Collaboration. In Section~2 we will calculate the various elementary
${\omega}N$ channels within a boson-exchange model and discuss the
uncertainties due to coupling constants and form factors.
The transport calculations within the HSD approach~\cite{Ehehalt}
for $\omega$-meson production in nucleus-nucleus reactions will be
presented in Section~3. A summary and
discussion of open questions concludes this work in Section~4.

\section{${\omega}N$ cross sections}
We start by recalling some general features of the
${\omega}N$ cross section. Since the $\omega$-meson mass
is about 5.5 times larger than the pion mass,
at low $\omega$-meson momenta the exothermic reactions
like ${\omega}N{\to}{\pi}N$, ${\omega}N{\to}2{\pi}N$,
${\omega}N{\to}3{\pi}N$   and other channels with a production
up to 5 $\pi$-mesons should be dominant  and \, the
corresponding inelastic \, cross section behave \, 
as \, $\sigma_{inel}{\simeq}1/p_\omega$, where \,  $p_\omega$
denotes the laboratory  momentum.
When the total mass of the final particle becomes equal
to the total mass of the incident ones, e.g. in the
${\omega}N{\to}{\rho}N$ or
${\omega}N{\to}{\omega}N$ reaction, the relevant cross section
should approach some  constant at low $p_\omega$.
On the other hand the endothermic inelastic reactions,
where the total mass of the produced
particles is larger than the initial mass, have a threshold
behaviour and are more important at high energies.

The amount of the available
${\omega}N$ reaction channels, especially
in the inelastic sector, is quite large. Thus one needs
a reliable model for the $\omega$-nucleon interaction at energies
up to few GeV. The complexibility of the problem has been addressed
in the recent studies from Klingl, Kaiser and
Weise~\cite{Klingl1,Klingl2} and Friman~\cite{Friman1,Friman2},
where some of the ${\omega}N$ reaction channels were
calculated. However, as pointed out in Refs.~\cite{Friman1,Friman2}
the calculated results are very sensitive to the
parameters of the model, i.e. coupling constants, form factors, etc.

\begin{figure}[t]
\psfig{file=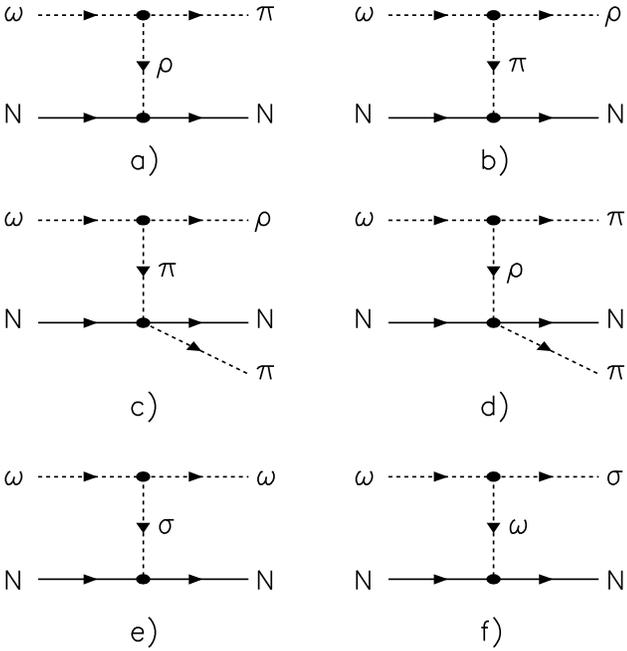,width=10cm}
\phantom{aa}\vspace{0.1cm}
\caption[]{The diagrams for the ${\omega}N$ reaction channels
evaluated in the text.}
\label{dia}
\end{figure}

As a first step for a simplification we adopt the
$\omega\rho\pi$ dominance model proposed by Gell-Mann,
Sharp and Wagner~\cite{GellMann} as well as the $\sigma$ exchange
approximation, which accounts for an effective two-pion exchange 
in the spin-isospin zero channel.
In practice we restrict to the diagrams shown in Fig~\ref{dia}
which contain the $\omega\rho\pi$-vertex and the
$\omega \sigma \omega$-vertex. In this way we expect to obtain
lower bounds especially on the total inelastic scattering cross section.

One way to construct the $\omega{\rho}\pi$ Lagrangian
is due to the current-field identities of Kroll, Lee and
Zumino~\cite{Kroll}, where the isoscalar and isovector parts of the
electromagnetic current -- with the $\omega$- and $\rho$-meson currents,
respectively -- can be identified. Starting with the Lagrangian
for the vector-meson photoproduction on the nucleon one can construct
the corresponding Lagrangian for the vector meson-nucleon
interaction as in Ref.~\cite{Friman}.
The photoproduction of $\rho$ and $\omega$-mesons on the nucleon
close to threshold ($E_\gamma{\leq}$2~GeV) has been analyzed
by Friman and Soyeur~\cite{Friman} also within the framework of
the one-boson exchange (OBE) model. In this sense the OBE
model can be applied
to elastic and inelastic ${\omega}N$ interactions. Furthermore,
the relevant coupling constant
$g_{\omega\rho\pi}$
as well as the corresponding form factor can be taken from the earlier
analysis in Ref.~\cite{Friman}.
In the present work we additionally consider
$\omega\sigma\omega$-vertices
and discuss the uncertainties related to the
coupling constants and form factors at the vertices. The latter
uncertainties are reduced to a large
extent by the analysis of experimental data on the
${\pi}N{\to}{\omega}N$ ( or inverse ${\omega}N{\to}{\pi}N$)
reaction.

\subsection{The channel ${\omega}N{\to}{\pi}N$}
The ${\omega}N{\to}{\pi}N$ cross section  can be controlled by the
experimental data on the inverse reaction via detailed balance.
The OBE diagram within our approach for the
$\pi{N}{\to}\omega{N}$ reaction is described by
the $\rho$-meson exchange as shown in Fig.~\ref{dia}a).
For our present study we adopt an effective Lagrangian for
the $\omega\rho\pi$ interaction as
\begin{equation}
{\cal L}_{\omega\rho\pi}=\frac{g_{\omega\rho\pi}}{m_{\omega}} \
\epsilon_{\alpha\beta\gamma\delta} \partial^\alpha\rho^\beta
\partial^\gamma\omega^\delta\pi  ,
\label{lag1}
\end{equation}
where the coupling constant $g_{\omega\rho\pi}=11.79$ was
evaluated from the ${\omega}{\to}3\pi$ partial width  assuming the virtual
$\rho$ model for the $\omega$-meson decay~\cite{GellMann}. Our result is
close to $g_{\omega\rho\pi}=10.88$ from
Ref.~\cite{Klingl}. In (\ref{lag1}) $\epsilon_{\alpha\beta\gamma\delta}$
denotes the antisymmetric tensor while  $\rho$, $\omega$ and $\pi$
are the corresponding meson fields. The ${\rho}NN$ Lagrangian is
taken as
\begin{eqnarray}
{\cal L}_{{\rho}NN}= - g_{{\rho}NN}
\left( \bar{N} \gamma^\mu \tau N \cdot \rho_\mu
\right. \nonumber \\
\left. + \frac{\kappa}{2 m_N} \bar{N} \sigma^{\mu \nu}
\tau N \cdot \partial_\mu \rho_\nu \right),
\label{rnn}
\end{eqnarray}
where $N$ stands for the nucleon field, $\tau$ for the
Pauli matrices, $g_{{\rho}NN}{=}3.24$ according to
Ref.~\cite{Sibirtsev1},
while the tensor coupling constant is given by the ratio
$\kappa{=}f_{{\rho}NN}/g_{{\rho}NN}{=}6.1$ .

\begin{figure}[t]
\psfig{file=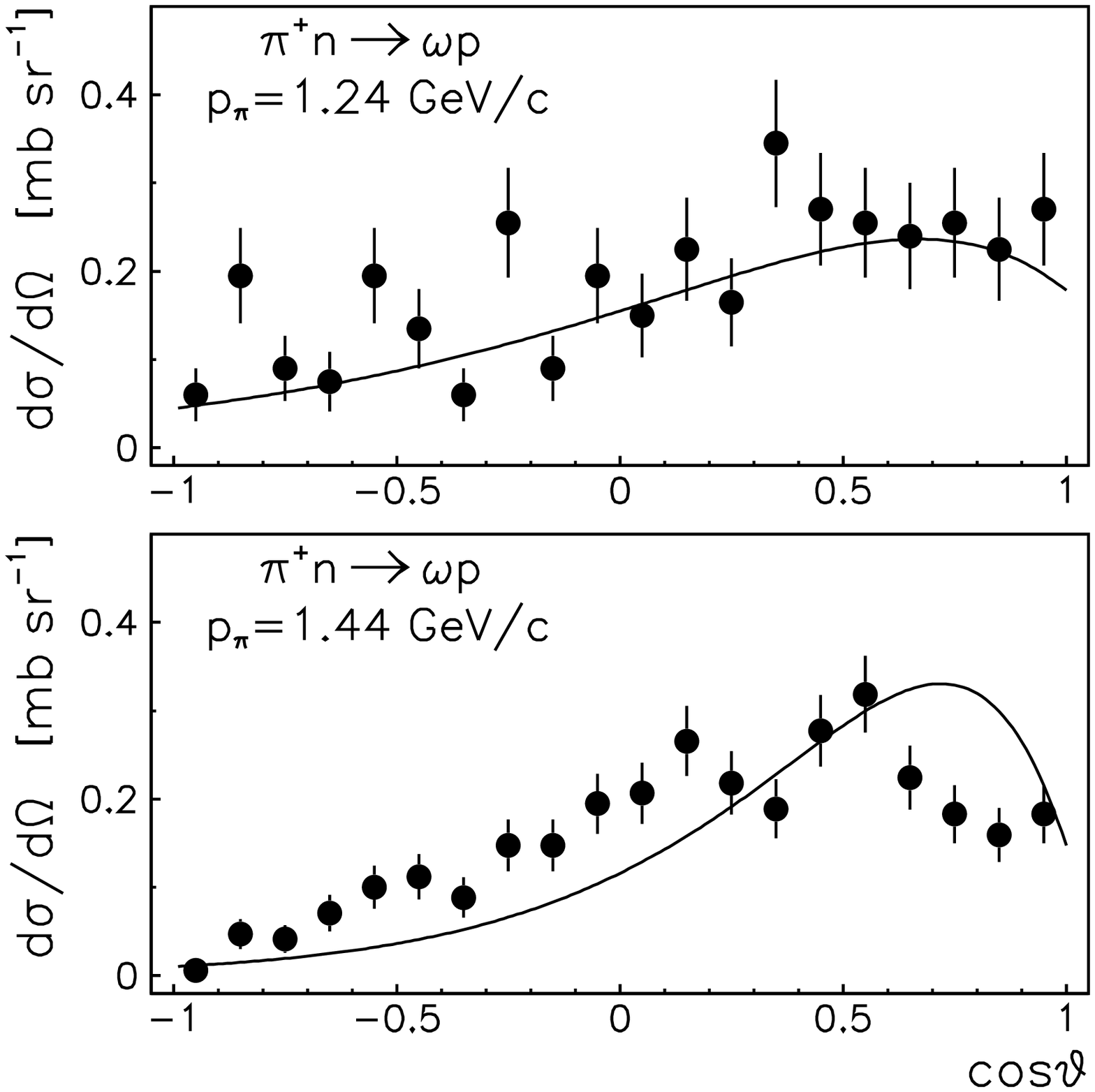,width=9.2cm,height=9cm}
\caption[]{Differential cross sections for the $\pi^+n{\to}{\omega}p$
reaction at pion momenta of 1.24 GeV/c and 1.44 GeV/c.
The data were taken from Ref.~\protect\cite{Danburg}
and are shown for 100~MeV-wide c.m. energy intervals while
the solid lines show our calculations.}
\label{om3ad}
\end{figure}

The differential cross section for the ${\pi}N{\to}{\omega}N$ process
is given as~\cite{Jackson}:
\begin{eqnarray}
\label{sigmao}
\frac{d\sigma}{dt} =
\frac{g^2_{\omega\rho\pi}}{m^2_\omega} \
\frac{1}{8\pi\lambda(s,m^2_N,m_{\pi}^2)} \
\frac{F^2_{\omega\rho\pi}F^2_{\rho N N}}
{(t-m^2_\rho)^2}  \nonumber   \\
\times \left\lbrack -\left(g_{\rho NN} + f_{\rho NN}\right)^2
m^2_{\omega}q^2_{\omega}t +
\left(g^2_{\rho NN} - \frac{f^2_{\rho NN}t}{4m^2_N}\right)
\right. \nonumber \\
\times \left. \left\lbrace
\frac{sin^2\theta}{8 \, s} \ \lambda(s,m^2_N,m_{\pi}^2) \
\lambda(s,m^2_N,m_{\omega}^2)
\right\rbrace
\right\rbrack
\end{eqnarray}
where
\begin{equation}
q^2_\omega=\frac{\lambda(t,m^2_{\omega},m^2_{\pi}) }{4m^2_\omega} .
\end{equation}
In Eq. (\ref{sigmao}) $s$ is the squared invariant collision energy,
$t$ is the 4-momentum
transfered from the initial to the final nucleon, $\theta$
is the production angle and
\begin{equation}
\lambda (x,y,z) = (x-y-z)^2-4yz.
\end{equation}
For the $\omega\rho\pi$ vertex we use the form factor
\begin{equation}
F(t) = \frac{\Lambda^2-m_\rho^2}{\Lambda^2-t} .
\label{fof}
\end{equation}

We note that using a formfactor similar to (\ref{fof}) for the
${\rho}NN$-vertex the energy dependence of the
${\pi}N{\to}{\omega}N$ production cross section as well as the
differential cross sections cannot be described
properly. Similar difficulties in comparing the $\rho$- exchange
model with experimental data were found by other
authors~\cite{Miller,Bacon,Cohn,Danburg,Klinglp}.
We, therefore, introduce a more general formfactor
at the ${\rho}NN$-vertex as\footnote{For a discussion
of the ansatz (\ref{form2}) see
Ref.~\cite{Gottfried} and references therein.}
\begin{equation}
F(t,s) = \exp{(\beta \, t)} \ \exp{(-\alpha \, s)}.
\label{form2}
\end{equation}

To fix the cut-off $\Lambda$ in the $\omega\rho\pi$ vertex as well as
the parameters $\beta$ and $\alpha$ in the ${\rho}NN$ vertex we fit
the available experimental  data on differential and total
cross sections for the ${\pi}N{\to}{\omega}N$ reaction.

\begin{figure}[h]
\phantom{aa}\vspace{-0.7cm}
\psfig{file=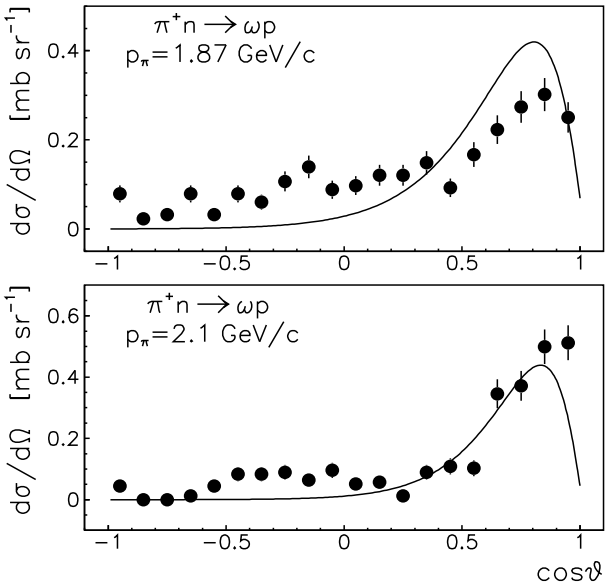,width=9.2cm,height=9cm}
\caption[]{Differential cross sections for the $\pi^+n{\to}{\omega}p$
reaction at pion momenta of 1.87 GeV/c and 2.1 GeV/c.
The data were taken from Ref.~\protect\cite{Danburg}
and are shown for 100~MeV-wide c.m. energy intervals while
the solid lines show our calculations.}
\label{om3bd}
\end{figure}

Figs.~\ref{om3ad},\ref{om3bd} show the $\pi^+n{\to}{\omega}p$
differential cross section~\cite{Danburg} for
pion  momenta from $p_\pi$=1.24 to 2.1~GeV/c
together with our calculations (solid lines). The data are shown for
100~MeV-wide $\sqrt{s}$ energy intervals centered at
the pion momenta indicated in Figs.~\ref{om3ad},\ref{om3bd}.
Note, that only  statistical errors are shown.

\begin{figure}[t]
\psfig{file=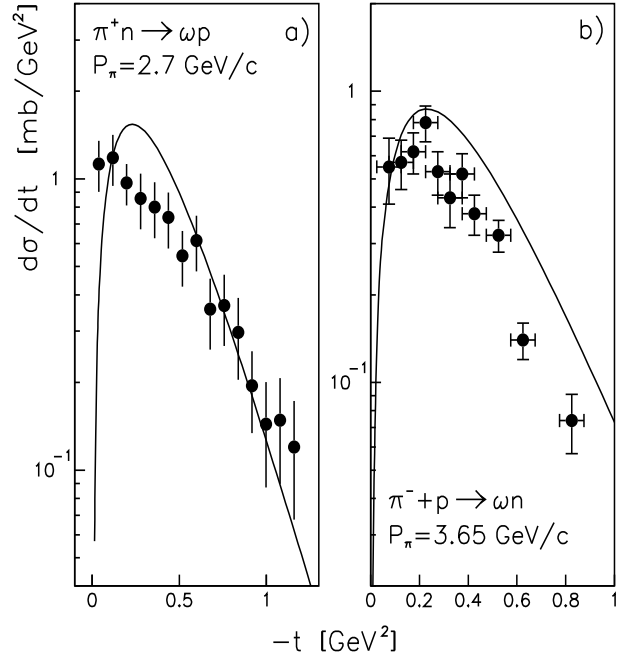,width=9.2cm,height=10cm}
\caption[]{The $\pi^+n{\to}{\omega}p$ (a) and $\pi^-p{\to}{\omega}n$
(b) differential cross sections as a function of the transverse
4-momentum squared. The data were taken from
Refs.~\protect\cite{Miller,Holloway} while the lines show our
calculations.}
\label{om3}
\end{figure}

Fig.~\ref{om3}a) shows the $\pi^+n{\to}{\omega}p$ differential
cross section~\cite{Miller} at the pion momentum $p_\pi{=}2.7$~GeV/c
together with our calculations while Fig.~\ref{om3}b) shows the
comparison to the $\pi^-p{\to}{\omega}n$ data~\cite{Holloway} at
$p_\pi{=}3.65$~GeV/c.

  From the fit of the experimental data with respect to the
$s$-dependence of the cross section and the differential
cross sections $d\sigma{/}dt$
the parameters  were fixed in the following way:
\begin{equation}
\Lambda=2.7 \ GeV ;\ \ \ \
\beta=2.3 \ GeV^{-2}; \ \ \ \
\alpha=0.16 \ GeV^{-2}.
\end{equation}

\begin{figure}
\psfig{file=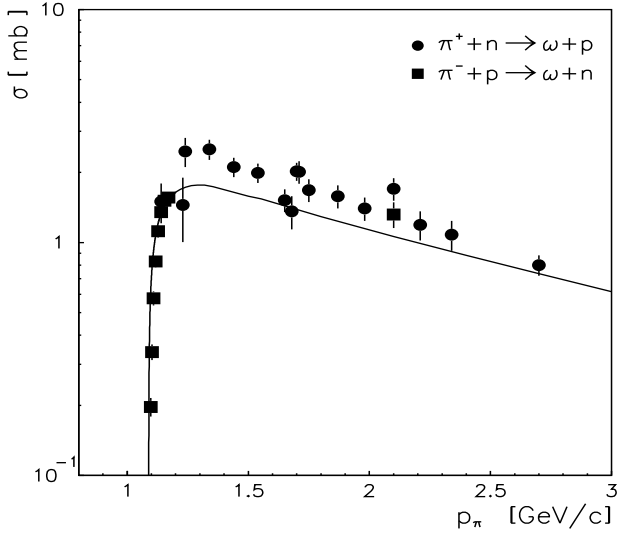,width=9.2cm,height=8cm}
\caption[]{The $\pi^+n{\to}{\omega}p$ (circles) and 
$\pi^-p{\to}{\omega}n$ (squares)
total cross sections as a function of the pion momentum. The data
are from Ref.~\protect\cite{LB} while the solid line shows 
our calculations for the cut-off parameter $\Lambda{=}2.7$~GeV.}
\label{om1}
\end{figure}

Fig.~\ref{om1} shows the total $\pi^+n{\to}{\omega}p$ and
$\pi^-p{\to}{\omega}n$ cross section as a function of the
pion momentum. Again the data are reasonably reproduced
with $\Lambda{=}2.7$~GeV, which we now fix for the following
analysis.

\begin{figure}
\psfig{file=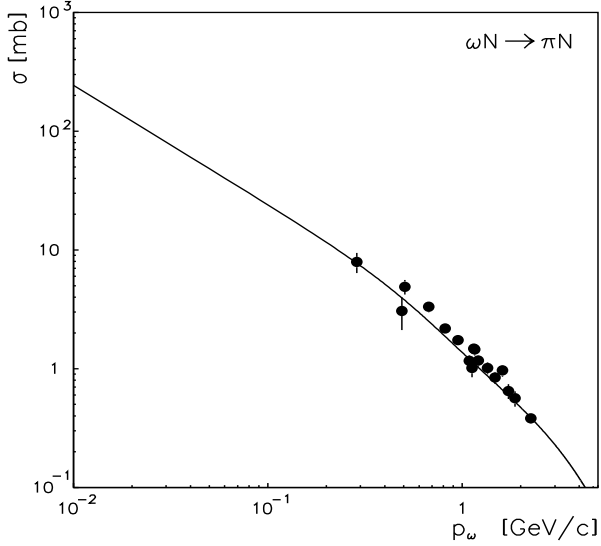,width=9.0cm,height=8cm}
\phantom{aa}\vspace{-0.1cm}
\caption[]{The cross section for the ${\omega}N{\to}{\pi}N$
reaction as a function of the laboratory $\omega$-momentum.
The solid line is our calculation while the full circles show the
${\omega}p{\to}\pi^+n$ experimental data
obtained by detailed balance.}
\label{om4}
\end{figure}

The calculated cross section for the inverse reaction
${\omega}N \to {\pi}N$ is shown in
Fig.~\ref{om4} for $\Lambda_1{=}2.7$~GeV as a function of the
$\omega$-momentum in the laboratory in comparison to the
experimental data obtained by detailed balance.

\subsection{The channel ${\omega}N{\to}{\rho}N$}
The ${\omega}N{\to}{\rho}N$ cross section can be calculated in the
one-pion exchange model using the diagram in Fig.~\ref{dia}b).
The Lagrangian for the $\omega\rho\pi$ interaction is given
by Eq.~(\ref{lag1}), while the ${\pi}NN$ Lagrangian is taken as
\begin{equation}
{\cal L}_{{\pi}NN}=
-ig_{{\pi}NN} \bar{N} \gamma_5 \tau N \cdot \pi
\label{lag3}
\end{equation}
with the coupling constant $g_{{\pi}NN}=$13.59~\cite{DeSwart}.

\begin{figure}
\psfig{file=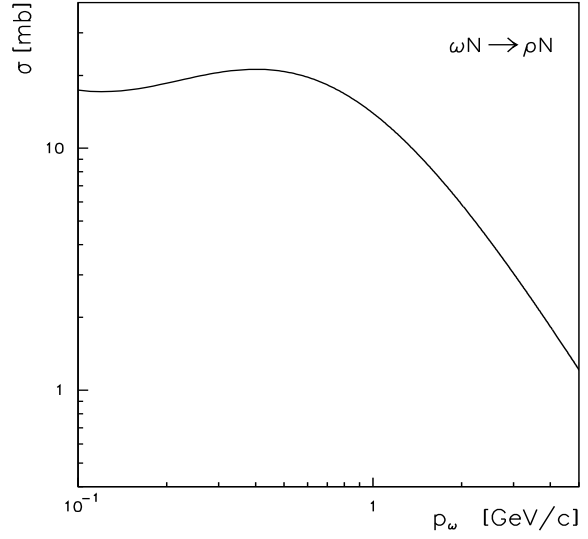,width=9.cm,height=8cm}
\caption[]{The cross section for the ${\omega}N{\to}{\rho}N$
reaction.}
\label{om4a}
\end{figure}

The ${\omega}N{\to}{\rho}N$ differential
cross section then is given by
\begin{eqnarray}
\frac{d\sigma}{dt}  =
-g^2_{{\pi}NN} \ t \ \frac{g^2_{\omega\rho\pi}}{m^2_\omega} \
\frac{(t-m^2_\omega-m^2_\rho)^2 - 4m^2_\omega m^2_\rho}
{96 \, \pi \ \lambda(s,m^2_\omega,m^2_N)} \nonumber \\
 \times  \frac{F^2_{\omega\rho\pi}(t) \ F^2_{{\pi}NN}(t)}
{(t-m^2_\pi)^2},
 \label{eq2}
\end{eqnarray}
where the coupling constant and the form factor for the
$\omega\rho\pi$-vertex is fixed by the ${\pi}N{\to}{\omega}N$
calculations.
The ${\pi}NN$ form factor was taken as in Eq.~(\ref{fof}) with the
cut-off parameter $\Lambda = 1.05$~GeV in line with 
Ref.~\cite{Tsushima}.

The resulting ${\omega}N{\to}{\rho}N$ cross section is shown by the
solid line in Fig.~\ref{om4a} as a function of the $\omega$-meson
laboratory momentum; the cross section is practically
constant at low momentum ($\approx$ 18 mb) and levels off at higher
momenta due to the form factors involved. However, above momenta of about
200 MeV/c it is already larger than for the ${\pi}N$ final channel.

\subsection{The channel ${\omega}N{\to}\rho{\pi}N$}
The relevant diagram for the reaction ${\omega}N{\to}\rho{\pi}N$
is shown in Fig.~\ref{dia}c) and the corresponding total cross
section can be calculated by using the Berestetsky-Pomeranchuk
approach~\cite{B-P} as
\begin{eqnarray}
\sigma  =  \frac{1}{48\pi^2 \lambda(s,m^2_N,m^2_\omega)} \!\!
\intop_{(m_N+m_\pi)^2}^{(\sqrt{s}-m_\rho)^2}\!\! ds_1
\lambda^{1/2}(s_1,m^2_\pi,m^2_N) \nonumber \\
 \times \sigma_{\pi N}(s_1)
\intop_{t_-}^{t^+} dt \ \frac{ g^2_{\omega\rho\pi} \
F^2_{\omega\rho\pi}}{m^2_\omega \ (t-m^2_\pi)^2}
\nonumber \\
\times  \left\lbrack  \left( t-m^2_\omega-m^2_\rho \right)^2
- 4m^2_\omega m^2_\rho \right\rbrack .
\label{mul1}
\end{eqnarray}

In Eq.(\ref{mul1})  $s_1$ is the squared invariant mass of the
final ${\pi}N$
system, $\sigma_{{\pi}N}$ is the ${\pi}N{\to}{\pi}N$ elastic
cross section taken from Ref.~\cite{LB} and
\begin{eqnarray}
t^\pm=m^2_\omega+m^2_\rho-\frac{1}{2s}
\left\lbrack (s+m^2_\omega-m^2_N) \right. \nonumber \\
\times (s+m^2_\rho-s_1)
\left. \mp \lambda^{1/2}(s,m^2_\omega,m^2_N) \,
\lambda^{1/2}(s,m^2_\rho,s_1)\right\rbrack .
\end{eqnarray}

\begin{figure}[h]
\phantom{aa}\vspace{-0.9cm}
\psfig{file=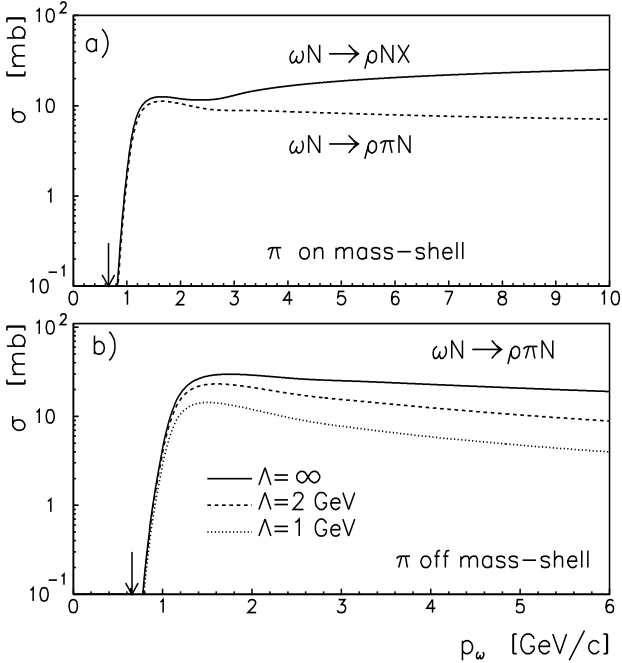,width=9.2cm,height=10cm}
\caption[]{a): The exclusive ${\omega}N{\to}\rho{\pi}N$
(dashed line) and inclusive ${\omega}N{\to}{\rho}NX$
cross sections calculated with the one-pion exchange model
and within the on mass-shell approximation. The arrow indicates
the threshold for the exclusive reaction. b): The exclusive
${\omega}N{\to}\rho{\pi}N$ cross section calculated within the
off mass-shell approach for different cut-off
parameters $\Lambda$ in the ${\pi}N{\to}{\pi}N$ vertex.}
\label{om6}
\end{figure}

The resulting ${\omega}N{\to}\rho{\pi}N$ cross section is shown in
Fig.~\ref{om6}a)  by the dashed line while the arrow
indicates the corresponding reaction threshold. Now replacing the
elastic ${\pi}N{\to}{\pi}N$ cross section in Eq.~(\ref{mul1})
by the total cross section $\pi N \rightarrow N X$
we obtain an estimate for the inclusive
${\omega}N{\to}{\rho}NX$ cross section, which is shown in terms of the
solid line in Fig.~\ref{om6}a). This channel provides an inelastic
${\omega}N$ cross section of about $\simeq$25 mb at higher momenta.

In Fig.~\ref{om6}a) we have adopted the on-shell
approximation~\cite{B-P}. The off-shell correction to the
${\pi}N$ amplitude can be estimated e.g. as proposed by
Ferrari and Selleri~\cite{Ferrari}, i.e. replacing $t{\to}m^2_\pi$
in the $\lambda$-function of Eq.~(\ref{mul1}) and
introducing the form factor
\begin{equation}
F(t) = \frac{\Lambda^2-m_\pi^2}{\Lambda^2-t}
\label{pifo}
\end{equation}
at the ${\pi}N{\to}{\pi}N$ vertex.

The solid line in Fig.~\ref{om6}b)
indicates the corresponding calculation when neglecting the form factor
at the ${\pi}N{\to}{\pi}N$ vertex ($\Lambda = \infty$),
the dashed line shows the
calculation with the cut-off parameter $\Lambda$=2~GeV
while the dotted line corresponds to $\Lambda$=1~GeV. Note,
that the uncertainty due to the cut-off parameter is quite
large when varying $\Lambda$ from 1 GeV to $\infty$.
However, the comparison between the  dashed line in
Fig.~\ref{om6}a) and the dotted line in Fig.~\ref{om6}b)
indicates that the results obtained for $\Lambda$=1~GeV (in line
with Ref. \cite{Engel})  are in
 reasonable agreement with the on mass-shell calculations.
Thus in the following we will adopt the on mass-shell approximation.

\subsection{The channel ${\omega}N{\to}2{\pi}N$}
The ${\omega}N{\to}2{\pi}N$ cross section can be calculated in the
$\rho$-meson exchange model as shown by the diagram in
Fig.~\ref{dia}d). By taking into account the on-shell
${\rho}N{\to}{\pi}N$ amplitude the total cross  section is given as
\begin{eqnarray}
\sigma =\frac{1}{32\pi^2 \ \lambda(s,m^2_N,m^2_\omega)} \!\!
\intop_{(m_N+m_\pi)^2}^{(\sqrt{s}-m_\pi)^2}\!\!\!\!\!\!ds_1
\lambda^{1/2}(s_1,m^2_\rho,m^2_N) \nonumber \\
\times \sigma_{{\rho}N{\to}{\pi}N}(s_1)
\intop_{t_-}^{t^+} dt \ \frac{ g^2_{\omega\rho\pi} \
F^2_{\omega\rho\pi}}{m^2_\omega \ (t-m^2_\rho)^2} \nonumber \\
\times \left\lbrack  \left( t+m^2_\omega-m^2_\pi \right)^2
- 4 t m^2_\omega \right\rbrack  ,
\label{mul2}
\end{eqnarray}
where
\begin{eqnarray}
t^\pm=m^2_\omega+m^2_\pi-\frac{1}{2s}
\left\lbrack (s+m^2_\omega-m^2_N)
\right. \nonumber \\
\times (s+m^2_\pi-s_1)
\left. \mp \lambda^{1/2}(s,m^2_\omega,m^2_N) \,
\lambda^{1/2}(s,m^2_\pi,s_1)\right\rbrack .
\end{eqnarray}

\begin{figure}[t]
\psfig{file=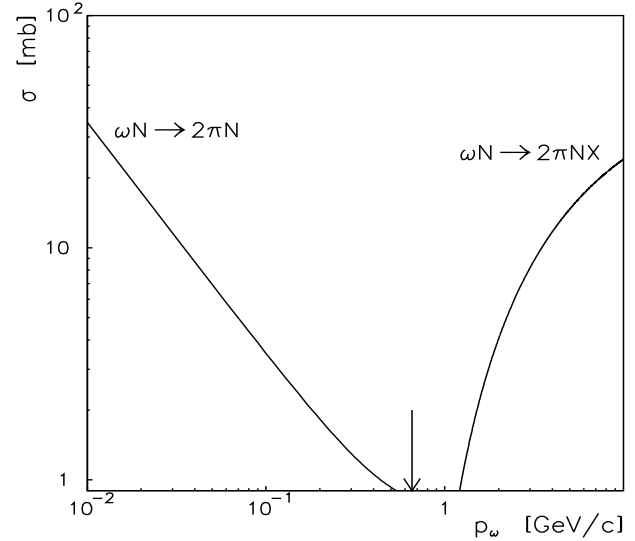,width=9.2cm,height=8cm}
\caption[]{The exclusive ${\omega}N{\to}2{\pi}N$
and inclusive ${\omega}N{\to}{\pi}NX$
cross sections. The arrow indicates the ${\omega}N{\to}\pi{\rho}N$
threshold using the $\rho$ pole mass.
The inclusive cross section is calculated for the
invariant mass of the ${\pi}N$-system above $m_\rho{+}m_N$.}
\label{om7}
\end{figure}

In Eq.~(\ref{mul2})  the ${\rho}N{\to}{\pi}N$ cross section
was taken as in Ref.~\cite{Sibirtsev2},
\begin{eqnarray}
\sigma_{{\rho}N{\to}{\pi}N}(s_1) = \frac{\pi^2}{3} \ \frac{a}
{\lambda^{1/2}(s_1,m^2_\rho,m^2_N)} \nonumber \\
\times \frac{\Gamma^2}{(\sqrt{s_1}-M)^2+\Gamma^2/4},
\label{pirho}
\end{eqnarray}
with the parameters $a$=413~$\mu$b/GeV$^2$, $M$=1.809~GeV and
$\Gamma$ = 0.99~GeV.
Note, that the function $\lambda^{1/2}$ in Eq.~(\ref{mul2})
is cancelled by Eq.~(\ref{pirho}).
The cross section ${\omega}N{\to}2{\pi}N$ calculated with
Eq.~(\ref{mul2}) is shown in Fig.~\ref{om7} and approaches 30~mb for
omega momenta of 10 MeV/c,  but dropps off fast with higher momentum.

Now we replace the ${\rho}N{\to}{\pi}N$ cross section in
Eq.~(\ref{mul2}) by the total ${\rho}N$ cross section
taken from Ref.~\cite{Kondratuyk} in order to estimate the
inclusive ${\omega}N{\to}{\pi}NX$ reaction due to
$\rho$-meson exchange. However,
we replace the lower limit of the first integral
in Eq.~(\ref{mul2}) by $(m_N+m_\rho )^2$, which allows
us to estimate only part of the total cross section.
This calculated cross section is shown in Fig.~\ref{om7}
together with the threshold given by the bare $\rho$-mass (arrow) for
momenta above 1 GeV/c.

\subsection{Elastic ${\omega}N{\to}{\omega}N$ scattering}
We describe elastic ${\omega}N$ scattering by the sigma-exchange
model shown in terms of the diagram Fig.~\ref{dia}e), which effectively
accounts for a correlated two-pion exchange in the spin-isospin zero channel. 
The Lagrangian used reads
\begin{eqnarray}
{\cal L}_{{\sigma}NN} =
g_{{\sigma}NN} \bar{N} N \cdot \sigma, \\
{\cal L}_{\omega\sigma\omega} =
g_{\omega\sigma\omega} \left( \partial^\alpha\omega^\beta
\partial_\alpha\omega_\beta -\partial^\alpha\omega^\beta
\partial_\beta\omega_\alpha \right) \sigma  ,
\label{lag4}
\end{eqnarray}
with the scalar coupling constant
$g_{{\sigma}NN}$=10.54  and monopole form
factors with cut-off parameter $\Lambda$=2~GeV
at the ${\sigma}NN$ vertex~\cite{Machleid}.

An upper limit for the $\omega\sigma\omega$ coupling can
be obtained from the
$\omega{\to}2\pi^0\gamma$ partial width assuming that
this decay entirely proceeds through the
$\omega{\to}\sigma\omega$ process followed by
the $\omega{\to}\gamma$ transition due to the
vector dominance model and the $\sigma{\to}2\pi^0$ decay.
Starting from the Lagrangian (\ref{lag4}) and integrating
over the $\sigma$ spectral function Post~\cite{Post} obtained
$g_{\omega\sigma\omega} \approx $ 5.7
taking into account the most recent data~\cite{PDG2} on the
$\omega{\to}2\pi^0\gamma$ decay.
A coupling $g_{\omega\sigma\omega}$=0.5 was
estimated in Refs.~\cite{Nakayama,Hanhart} from
the $\omega{\to}2\pi^0\gamma$ decay~\cite{PDG0}
using, however, a different Lagrangian for the $\omega\sigma\omega$
interaction and replacing the $\sigma$-meson spectral function by a
$\delta$-distribution.

\begin{figure}[h]
\phantom{aa}\vspace{-0.7cm}
\psfig{file=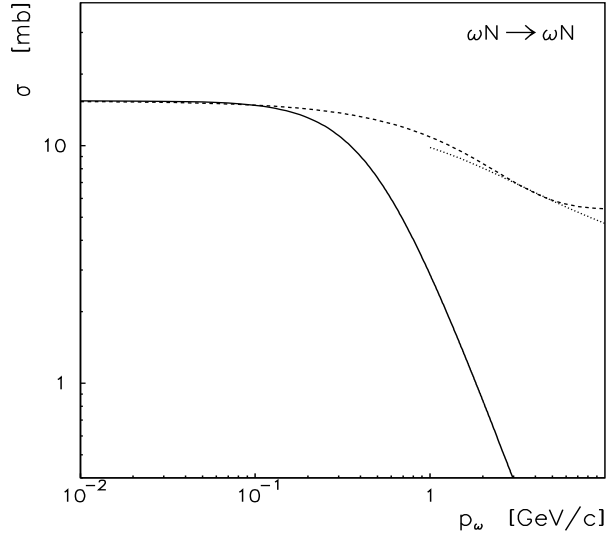,width=9cm,height=8cm}
\caption[]{The ${\omega}N{\to}{\omega}N$ elastic cross section.
The solid line indicates the calculation within the $\sigma$-exchange
model for $g_{\omega \sigma \omega}$=1.76; the dotted line is
the high energy limit~(\protect\ref{limit}), while the dashed line shows
our interpolation~(\protect\ref{par1}).}
\label{om5}
\end{figure}

On the other hand, as proposed by Singer~\cite{Singer},
the $\omega{\to}2\pi\gamma$ decay may proceed also through
the two-step process
$\omega{\to}\rho\pi$ followed by the $\rho{\to}\pi\gamma$
transition; the data on the partial $\omega \to 2\pi^0\gamma$
width indeed can be entirely described by the latter process.
A more sophisticated analysis has been performed by Fajfer and
Oakes~\cite{Fajfer}, who consider a three-step process
$\omega{\to}\rho\pi$ followed
by $\rho{\to}\pi\omega$ and as a last step the
$\omega{\to}\gamma$ transition. They predict \cite{Fajfer}
a branching ratio for the \, $\omega{\to}2\pi^0\gamma$ \,
decay of $8.21{\times}10^{-5}$, which is compatible to
$(7.2{\pm}2.5){\times}10^{-5}$ as quoted recently
by the PDG~\cite{PDG2}.

We point out that there are no reliable constraints
for the $\omega\sigma\omega$ coupling constant and
in the following calculations adopt
$g_{\omega\sigma\omega}$=1.76 keeping it essentially as a free
parameter of the model. To demonstrate the uncertainty, we will also
present final results for $g_{\omega\sigma\omega}$ = 0.
Furthermore, we use a monopole form factor
with cut-off parameter $\Lambda$=2.0~GeV for the $\omega\sigma\omega$
vertex. With the latter value  $g_{\omega\sigma\omega}$=1.76 the
resulting elastic and inelastic contributions to the
${\omega}N$ cross section will be quite moderate.

The elastic ${\omega}N{\to}{\omega}N$ cross
section then can be calculated as
\begin{eqnarray}
\frac{d\sigma}{dt} = \frac{g^2_{{\sigma}NN} \, g^2_{\omega\sigma\omega}}
{16\pi \,m^2_\omega \,  \lambda(s,m^2_\omega,m^2_N) }
\ \left( 4m^2_N - t \right) \nonumber \\
\frac{F^2_{\omega\sigma\omega} \,
F^2_{{\sigma}NN}} { (t-m^2_\sigma)^2} \
\left( m^4_\omega - \frac{m^2_\omega \, t }{3} + \frac{t^2}{12}
\right)
\end{eqnarray}
with $m_\sigma$=550~MeV. The result of this model is shown by the solid
line in Fig.~\ref{om5} and indicates a maximum of $\approx$15 mb at low
relative momentum but levels off very fast for high momenta.

The high energy limit for the ${\omega}N{\to}{\omega}N$
elastic cross section can be taken in the Quark
Model~\cite{Lipkin,Kajantie} as
\begin{equation}
\sigma_{{\omega}N{\to}{\omega}N}(s) = \frac{1}{2}
\left\lbrack \sigma_{\pi^+N{\to}\pi^+N}(s) +
\sigma_{\pi^-N{\to}\pi^-N}(s) \right\rbrack ,
\label{limit}
\end{equation}
which is shown in Fig.~\ref{om5} by the dotted
line above momenta of 1 GeV/c.
We add that in Eq. (\ref{limit}) the elastic $\pi^\pm{N}$
cross section is taken from Ref.~\cite{PDG1} and the
identity~(\ref{limit}) is used at the same invariant
energy for the ${\omega}N$- and ${\pi}N$-interaction.

For our following applications to heavy-ion transport simulations
we interpolate the ${\omega}N$ elastic cross section within the
range  from 10~MeV/c up to 10~GeV/c as
\begin{equation}
\label{par1}
\sigma_{el} = 5.4 + 10 \exp(-0.6 p_\omega) \ [{\rm mb}] ,
\end{equation}
where  $p_\omega$ denotes
the laboratory momentum of the $\omega$-meson in GeV/c. The
parameterization~(\ref{par1}) is shown additionally by the
dashed line in  Fig.~\ref{om5}.

\begin{figure}[t]
\psfig{file=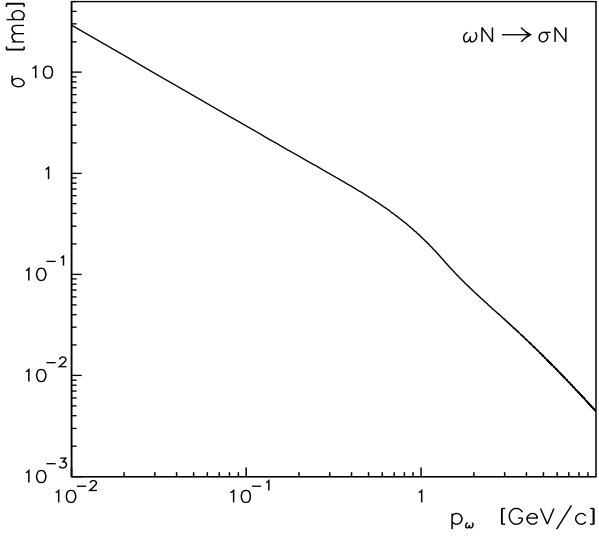,width=9.cm,height=8cm}
\caption[]{The ${\omega}N{\to}{\sigma}N$  cross section
within the $\omega$-exchange model for $g_{\omega\sigma\omega}$=1.76.}
\label{om12}
\end{figure}

\subsection{The channel ${\omega}N{\to}{\sigma}N$}
When incorporating
the $\omega\sigma\omega$-vertex for elastic $\omega N$ scattering one
has to consider this vertex also in the
${\omega}N{\to}{\sigma}N$ reaction due to
$\omega$-meson exchange as shown in Fig.~\ref{dia}f).
The differential cross section for this process is given by
\begin{eqnarray}
\frac{d\sigma}{dt} = \frac{g^2_{{\omega}NN} \, g^2_{\omega\sigma\omega}}
{96\pi \, m^2_\omega} \,
\frac{ 2t+ 4m^2_N}{\lambda(s,m^2_\omega,m^2_N)} \,
\frac{F^2_{\omega\sigma\omega} \,
F^2_{{\omega}NN}} { (t-m^2_\omega)^2} \nonumber \\
\left\lbrack(t-m^2_\omega-m^2_\sigma)^2 - 4m^2_\omega m^2_\sigma
\right\rbrack .
\end{eqnarray}
We introduce a monopole form factor at the ${\omega}NN$
vertex with the cut-off $\Lambda$=1.5~GeV~\cite{Machleid}.
The parameters for the $\omega\sigma\omega$ vertex
are the same as for the ${\omega}N{\to}{\omega}N$
calculations. This results in the ${\omega}N{\to}{\sigma}N$
cross section  shown in Fig.~\ref{om12}, which also gives up to 30 mb for
low $\omega$ momenta.

\subsection{The total ${\omega}N$ cross section}
\begin{figure}[b]
\phantom{aa}\vspace{-0.8cm}
\psfig{file=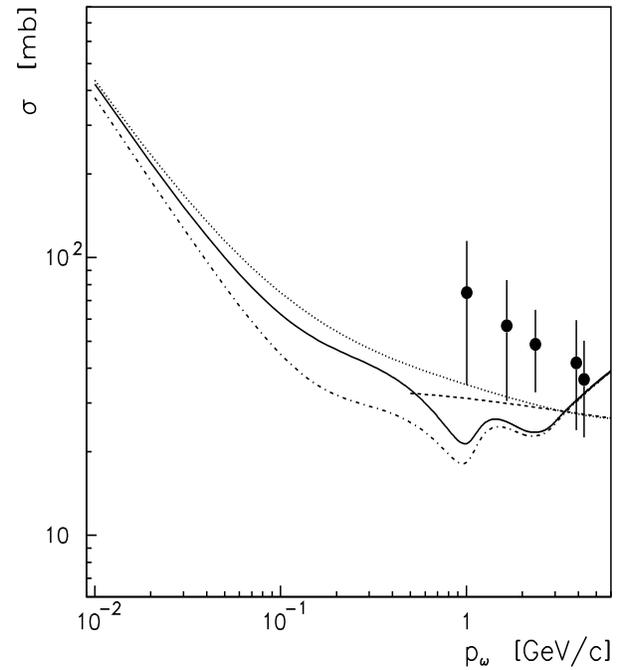,width=9cm,height=10cm}
\caption[]{The total ${\omega}N$ cross section as a function
of the $\omega$-meson momentum in the laboratory system.
The solid line shows our result as an incoherent sum of the
calculated partial cross sections; the dash-dotted line is obtained
when omitting the diagrams 1e) and 1f), i.e. for 
$g_{\omega\sigma\omega}$=0. The full circles show the
results obtained within the  VDM from the data on forward
$\omega$-photoproduction. The dashed line illustrates the
estimation from the Quark Model while the dotted line is
our actual parameterization.}
\label{om9}
\end{figure}

We finally incoherently sum the partial cross sections - calculated
for the different ${\omega}N$ reaction channels - and show
the result in terms of the solid line in Fig.~\ref{om9}. In order to
demonstrate the uncertainty due to the unknown coupling 
$g_{\omega\sigma\omega}$ we also show the total cross for 
$g_{\omega\sigma\omega}$=0 (dash-dotted line), i. e. neglecting 
the diagrams 1e) and 1f). In this limit the cross section is
on average reduced by about 25\%. 

In this summation we have taken the cross section
for the inclusive ${\omega}N{\to}{\rho}NX$ reaction
instead of the exclusive ${\omega}N \to \rho{\pi}N$ reaction.
Again at high momenta our result can be compared with the
estimate from the Quark Model~\cite{Lipkin,Kajantie}
obtained with Eq.~(\ref{limit}) by replacing the elastic
${\pi}N$ cross section with the total cross section. This
result is shown by the dashed line in Fig.~\ref{om9}
and matches with our calculations in the OBE model
at momenta of 0.5 GeV/c.

Furthermore, at high momenta the ${\omega}N$ total cross section
can be estimated within the Vector Dominance Model, where
the total ${\omega}N$ cross section can be related to
the forward ${\gamma}N\to \omega N$ differential cross section
as~\cite{Bauer}
\begin{equation}
\label{vector}
{\sigma}^2_{{\omega}N}(s) = {\frac{{\gamma}^2_{\omega}}{4\pi}}
{\frac{64 \pi}{\alpha}} \ \frac{1}{1+ \alpha_\omega^2} \
{\left(\frac{q_{\gamma}}{q_{\omega}} \right)^2}
{\left. \frac{d\sigma_{\gamma p {\to} \omega p}(s)}
{dt} \right|_{t=0} } ,
\end{equation}
where $q_{\gamma}$ and $q_{\omega}$ are the photon and
$\omega$-meson momenta in the ${\gamma}N$ and ${\omega}N$
center-of-mass systems at the same invariant
collision energy $\sqrt{s}$. In Eq. (\ref{vector}) $\alpha_\omega$
denotes the ratio of the real to imaginary part of the
${\omega}N$ forward scattering amplitude~\cite{Bauer},
while $g_\omega$ is the $\gamma\omega$ coupling constant.
We neglect $\alpha_\omega$ and take
$g^2_\omega/4\pi$=4.9 from Ref.~\cite{Ballam}.
The full circles (with error bars)
in Fig.~\ref{om9} show the total ${\omega}N$
cross section  obtained from the experimental data
on forward $\omega$-meson photoproduction~\cite{Ballam,LB1}.

\begin{figure}
\psfig{file=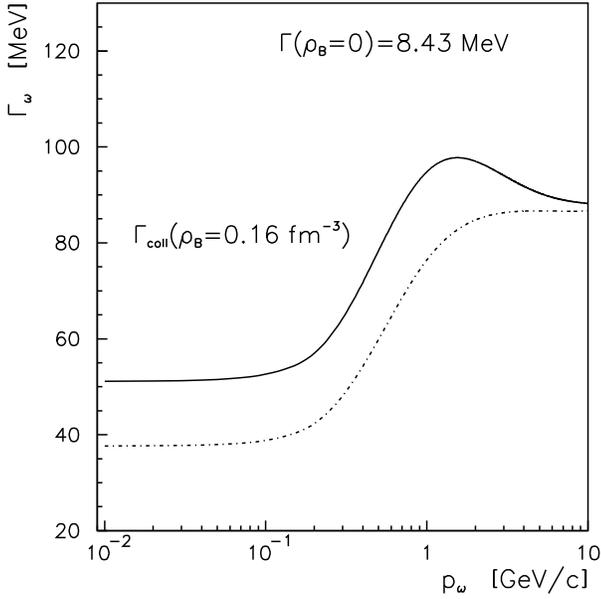,width=9.cm,height=9cm}
\caption[]{The in-medium $\omega$-meson width due to
collisional broadening as a function of the laboratory
$\omega$-meson momentum. The solid line is obtained when including all
channels while the dash-dotted line results for 
$g_{\omega\sigma\omega}$=0, i.e. neglecting diagrams 1e) and 1f).}
\label{om10}
\end{figure}

For the following transport calculations the total inelastic
${\omega}N$ cross section is separately fitted and interpolated as
\begin{equation}
\sigma_{inel} = 20 + \frac{4.0}{p_\omega} \ [{\rm mb}]    ,
\label{par2}
\end{equation}
where the laboratory $\omega$-meson
momentum $p_\omega$ is given in $GeV/c$. The parameterization for
the total ${\omega}N$ cross section then is given as a sum of the
elastic~(\ref{par1}) and  inelastic~(\ref{par2}) cross section
as shown by the dotted line in Fig.~\ref{om9}. Especially the inelastic
$\omega N$ cross section is found to be quite large at low momenta and
implies substantial final state interactions of the $\omega$-meson in
the medium. This also holds for the limit $g_{\omega\sigma\omega}$=0.

In line with these final state interactions the $\omega$-meson
will change its spectral function in the medium and (in first order)
will acquire a larger width due to collisional broadening at
baryon density $\rho_B$,
\begin{equation}
\Gamma_{coll}(p_\omega) = \frac{4}{(2 \pi)^3}
\int d^3p \ v_{{\omega}N}\  \sigma_{{\omega}N} (\sqrt{s})
\ \Theta(p_F - |{\bf p}|),
\label{colli}
\end{equation}
with $v_{{\omega}N}$ denoting the relative velocity
between the $\omega$-meson and the nucleon; $p_F$ is the nuclear
Fermi momentum and $\sqrt{s}$ the invariant energy of the
${\omega}N$ system.

The result for $\Gamma_{coll}$
at density $\rho_0$ ($p_F{\approx}0.26$~GeV/c) is shown
in Fig.~\ref{om10} by the solid line as a function of the
$\omega$-meson momentum relative to the nuclear
medium and is about 50 MeV at low
$\omega$ momenta, but increases to 100 MeV
above $p_\omega{\approx}$1 GeV/c. In the limit 
$g_{\omega\sigma\omega}$ = 0
(dash-dotted line) the collisional width is about 25\% smaller 
up to relative momenta of 1 GeV/c; this represents the lower limit in our
present approach. 
Our result is substantially larger at $p_\omega \approx$ 0 than
the estimate of 20 MeV in Ref. \cite{Friman2} and slightly larger than the
result of Klingl et al.~\cite{Klingl4} due to the additional channels
taken into account in our computations. Nevertheless,
the collisional width
for $\omega$-mesons produced at rest in nuclei is still small compared to
its mass.

\section{$\omega$-production in heavy-ion collisions}
The production of $\omega$-mesons from heavy-ion collisions
has been calculated previously within the HSD transport
approach~\cite{Bratkovskaya} and is described e.g. in
more detail in Ref.~\cite{PREP}.

\begin{figure}
\psfig{file=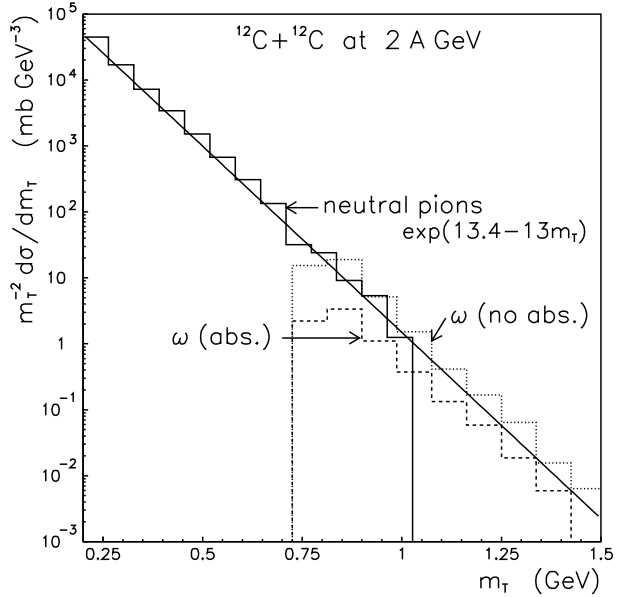,width=9cm}
\caption[]{The inclusive transverse mass spectra for  $\pi^0$
(solid histogram) and $\omega$-mesons (divided by a
factor of 3 due to the $\omega$ polarizations) from
$C{+}C$ collisions at 2~A$\cdot$GeV.
The dashed histogram
shows the calculations with all ${\omega}N$ interactions
while the dotted histogram is obtained without $\omega$ absorption.
The thick solid line indicates the $m_T$-scaling.}
\label{ombuu1}
\end{figure}

In these transport simulations the $\omega$-absorption as well as
elastic scattering in the
nuclear environment are explicitely included as well as (optional)
a reduction of the $\omega$ pole mass at finite baryon density.
In the present study we concentrate on the impact of
the ${\omega}N$ final state interactions with nucleons
especially with respect to a global  $m_T$-scaling
suggested by Bratkovskaya et al.~\cite{Bratkovskaya}
for  meson production from heavy-ion collisions at SIS energies.
The latter scaling was found to be well in line with the spectra
for $\pi^0$ and $\eta$ mesons as measured by the TAPS
Collaboration~\cite{TAPS1,TAPS2}.

Employing the elastic and inelastic $\omega N$ cross
sections from Section~2 we have performed calculations without
$\omega$-potentials in the medium
for the systems $^{12}C{+}^{12}C$, $^{40}Ca{+}^{40}Ca$ at 2 A GeV and
$^{58}Ni{+}^{58}Ni$ at 1.9 A$\cdot$GeV, which are studied
experimentally by the TAPS Collaboration also for $\omega$-meson
production via the Dalitz decay $\omega{\to}\pi^0\gamma$.

\begin{figure}
\psfig{file=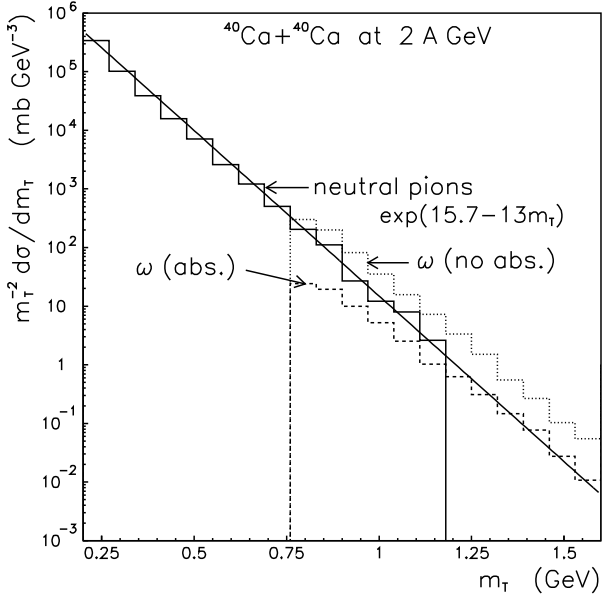,width=9.cm}
\caption[]{The inclusive  transverse mass spectra for  $\pi^0$
(solid histogram) and $\omega$-mesons
(divided by the number of polarizations) from $Ca{+}Ca$ collisions
at 2~A$\cdot$GeV.
The dashed histogram
shows the calculations with all ${\omega}N$ interactions
while the dotted histogram is obtained without $\omega$ absorption.
The thick solid line indicates the $m_T$-scaling.}
\label{ombuu2}
\end{figure}

\begin{figure}
\psfig{file=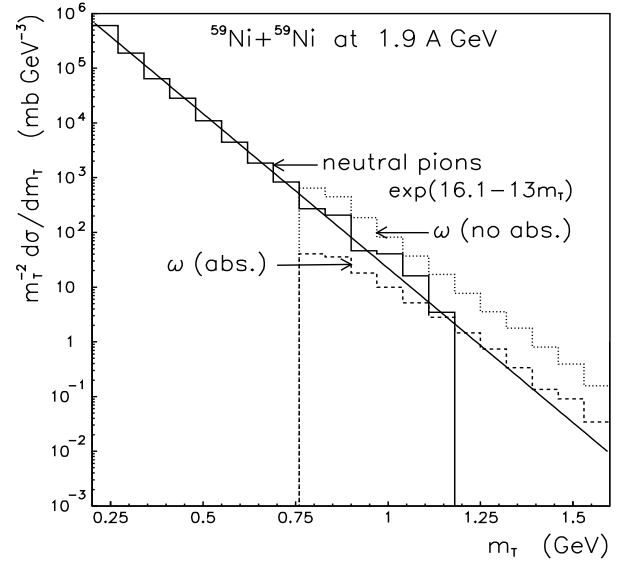,width=9cm,height=8.4cm}
\caption[]{The inclusive transverse mass spectra for  $\pi^0$
(solid histogram) and $\omega$-mesons from $Ni{+}Ni$ collisions
at 1.9~A$\cdot$GeV.
The dashed histogram
shows the calculations with all ${\omega}N$ interactions
while the dotted histogram is obtained without $\omega$ absorption.
The thick solid line indicates the $m_T$-scaling.}
\label{ombuu4}
\end{figure}

Figs.~\ref{ombuu1},\ref{ombuu2},\ref{ombuu4} show the calculated
inclusive transverse mass spectra for $\pi^0$ and $\omega$-mesons with
the transverse mass defined as $m_T=\sqrt{p_T^2+m^2}$,
where $p_T$ is the transverse momentum and $m$ stands
for the mass of the meson. The solid histograms in
Figs.~\ref{ombuu1},\ref{ombuu2},\ref{ombuu4} show the
$\pi^0$-meson spectra while the straight solid lines indicate
the scaling
\begin{equation}
\frac{1}{m_T^2} \ \frac{d\sigma}{dm_T} = A\, \exp(-Bm_T)
\label{scale}
\end{equation}
with parameters $A$ and $B$ given in Table~\ref{tab1}.

Indeed the neutral pions follow the scaling within the statistical
accuracy. The transverse mass distributions for the $\omega$-meson,
which were divided by a factor of 3 as in Ref.~\cite{Bratkovskaya}
due to the 3 different polarizations of the $\omega$-meson, are shown
by the dotted histograms when neglecting the $\omega$-meson absorption
in nuclear matter due to the inelastic ${\omega}N$
interactions. In this limit the $m_T$-scaling is overestimated
especially for $Ca{+}Ca$ and $Ni{+}Ni$.
The dashed histograms show the calculations accounting for the
$\omega$-absorption as well as ${\omega}N{\to}{\omega}N$
elastic scattering. In all cases the
$m_T$-scaling relative to neutral pions is underestimated especially for
low transverse mass of the $\omega$-mesons which results from the large
absorption cross section for $\omega$-mesons at low relative momenta
according to Section 2.

\begin{table}
\caption{\label{tab1}Parameters of the $m_T$-scaling
approximation~(\protect\ref{scale}).}
\begin{tabular}{lccc}
System & $C+C$  & $Ca+Ca$ & $Ni+Ni$ \\
Energy (A$\cdot$GeV) & 2.0 & 2.0 & 1.9 \\
$A$ (b$\cdot$GeV$^{-3}$) & 660 & 6583 & 9821 \\
$B$ (GeV$^{-1}$) & 13 & 13 & 13 \\
\end{tabular}
\end{table}
\vspace{0.5cm}

\begin{figure}[b]
\phantom{aa}\vspace{-0.9cm}
\psfig{file=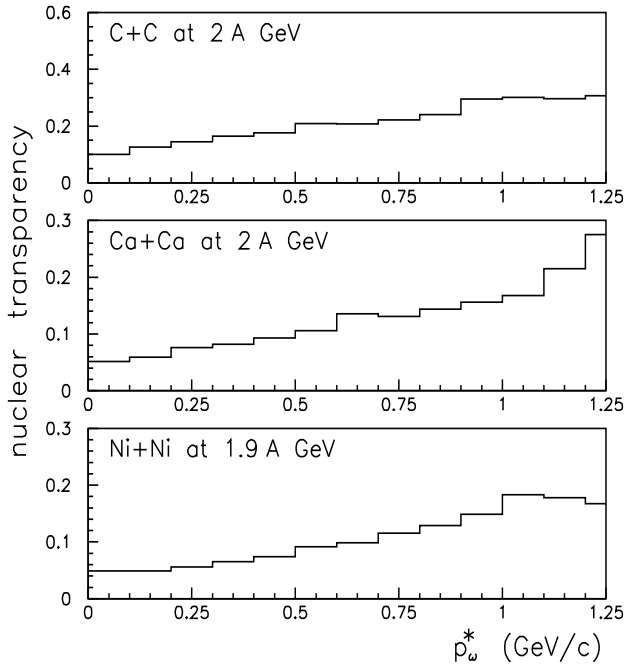,width=9.cm,height=10cm}
\caption[]{The nuclear transparency coefficient as a function of the
$\omega$-momentum in the center-of-mass system
for $C+C$, $Ca+Ca$ and $Ni+Ni$ collisions at 2.0 and 1.9 A GeV,
respectively.}
\label{ombuu3}
\end{figure}

Fig.~\ref{ombuu3} shows the nuclear transparency coefficient as a function
of the $\omega$-meson momentum in the center-of-mass
for $C+C$, $Ca+Ca$ and $Ni+Ni$ systems. Here the nuclear
transparency coefficient is defined as the ratio of the $\omega$-mesons
detected assymptotically for $t{\to}\infty$
to the total number of the $\omega$-mesons produced in
elementary pion-baryon and baryon-baryon collisions.
The effect is most pronounced for slow $\omega$-mesons
since the inelastic ${\omega}N$ interaction is very
strong at low momenta (cf. Fig.~\ref{om9}).

Table~\ref{tab2} shows the inclusive $\omega$-meson production cross
section for $C+C$, $Ca+Ca$ and $Ni+Ni$ collisions calculated
with and without ${\omega}N$ final state interactions.
Even for the light system $C+C$ the $\omega$ production
cross section is reduced by a factor $\simeq$ 4.5 due
to strong absorption.

\begin{table}
\caption{\label{tab2}Inclusive $\omega$-meson production
cross section from $C+C$, $Ca+Ca$ and $Ni+Ni$ collisions
calculated with and without final state interactions of the
$\omega$ mesons.}
\begin{tabular}{lccc}
System & $C+C$  & $Ca+Ca$ & $Ni+Ni$ \\
Energy (A$\cdot$GeV) & 2.0 & 2.0 & 1.9 \\
$\sigma$ (mb) with FSI & $1.42$ & $10.56$ & $19.5$ \\
$\sigma$ (mb) without FSI & $6.52$ & $91.4$ & $206.8$
\end{tabular}
\end{table}

\section{Summary}

Within the meson exchange model we have calculated different
partial cross sections for the ${\omega}N$ interaction, i.e.
${\omega}N{\to}{\pi}N$,
${\omega}N{\to}{\rho}N$, ${\omega}N{\to}{\omega}N$,
${\omega}N{\to}{\rho}{\pi}N$,
${\omega}N{\to}{\rho}NX$, ${\omega}N{\to}2{\pi}N$,
${\omega}N{\to}2{\pi}NX$ and  ${\omega}N{\to}{\sigma}N$
reaction channels. The free parameters of the model were fixed by
the available experimental data on the ${\pi}N{\to}{\omega}N$
reaction~\cite{Miller,Holloway,LB} or adopted from
the study on vector meson photoproduction~\cite{Friman} except for
the $\omega \sigma \omega$-vertex, where we have adopted a conservative
low coupling constant; lower limits for our cross sections are obtained 
for $g_{\omega\sigma\omega}$ = 0. 
We note that the OBE calculations performed here
do not correspond to {\it ab initio} calculations but serve as a
convenient method to extrapolate from available data to unknown, but
related channels. We estimate the relative error in the various cross
sections to be within a factor of 2.

At high energies, i.e. $\omega$ momenta above a few GeV/c,
the total cross section approaches a conventional hadronic cross section
of $\simeq$ 25~mb according to
the prediction from the Quark Model as well as the estimate from
the Vector Dominance model. The total cross section at low $\omega$
momenta according to the OBE model sums up to a
few hundred mb which indicates
a sizeable rescattering and reabsorption of the $\omega$-meson in the
nuclear medium.  The collisional
broadening of the $\omega$-meson at nuclear matter density
$\rho_0$ amounts to 40 - 50 MeV at rest, but increases with momentum
up to 80 - 100 MeV. Thus the $\omega$ spectral function at $\rho_0$ is
dominated by the hadronic couplings in the medium. This statement holds
also in view of the uncertainties still involved in the OBE approach.

Furthermore, we have investigated the impact of the $\omega N$
final state interactions on $\omega$-production
in heavy-ion collisions around 2 A$\cdot$GeV. It is found that
due to strong absorption the total $\omega$-meson
production cross section from $C+C$ collisions at SIS
energies is reduced by a factor of $\simeq$4.5 and
by a factor of $\simeq$10 for $Ni+Ni$ collisions.
Furthermore, we  find a significant deviation from the
$m_T$-scaling behaviour predicted in Ref.~\cite{Bratkovskaya} - where
lower $\omega N$ cross sections from Ref. \cite{Golub2} had been 
employed - 
for $\omega$-meson production in heavy-ion collisions due to the
strong final state interactions for slow $\omega$-mesons. These $m_T$
distributions might be controlled by the TAPS Collaboration in near
future and provide further experimental constraints on the 
${\omega}N$ interaction.

\acknowledgement{
The authors like to thank  V. Metag and  H. L\"ohner
for valuable discussions throughout this study. Furthermore, they like
to thank M. Post and G. Penner for pointing out a mistake in Eq. (3) in
an earlier version of this manuscript.
This work was supported by BMBF, Forschungszentrum J\"ulich
and the Russia Foundation for the fundamental Research.}

\end{document}